\documentclass{ws-mpla}
\providecommand{\Journal}[4]{{\it #1} {\bf #2}, #3 (#4)}
\providecommand{\NP}{Nucl. Phys. } %
\providecommand{\PR}{Phys. Rev. } %
\providecommand{\PRD}{Phys. Rev. D } %
\providecommand{\RMP}{Rev. Mod. Phys. } %
\providecommand{\RMP}{Rev. Mod. Phys. } %
\providecommand{\JPG}{J.Phys.G } %

\begin{document}

\markboth{Y.Chen, B.Q Ma} {Mixing of $1/2^-$ Octets under SU(3)
Symmetry}

\catchline{}{}{}{}{}

\title{Mixing of $1/2^-$ Octets under SU(3) Symmetry}

\author{Yan Chen}
\address{School of Physics, Peking University,
Beijing 100871, China}

\author{\footnotesize Bo-Qiang Ma}
\address{School of Physics and MOE Key Laboratory of Heavy Ion Physics,
Peking University, Beijing 100871, China\\
mabq@phy.pku.edu.cn}

\maketitle

\pub{}{Revised (Day Month Year)}

\begin{abstract}
We investigate the $J^p=1/2^-$ baryons in the octets based on flavor
$SU(3)$ symmetry. Since baryons with same quantum numbers can mix
with each other, we consider the mixing between two octets before
their mixing with the singlet. Most predicted decay widths are
consistent with the experimental data, and meanwhile we predict two
possible $\Xi$ mass ranges of the two octets.

\end{abstract}

\ccode{PACS Nos.: 14.20.-c, 11.30.Hv}

\section{Introduction}

The hypothesis of the approximate SU(3) symmetry of strong
interactions proposed by Gell-Mann and Ne'eman~\cite{Gell,Neeman,GZ}
has been proved quite successful and fruitful in the classification
of elementary particles. In this classification scheme, one can
group the experimentally known strongly interacting particles with
the same quantum numbers of spin and parity into various irreducible
representations of the SU(3) group~\cite{SGM}. The assignment of
baryons and mesons to definite SU(3) multiplets seems to be very
satisfactory for the $1/2^+$, $3/2^+$ baryons and $0^-$, $1^-$
mesons. It also seems to be satisfactory for the particles with
spin-parity $3/2^-$, $5/2^-$, $7/2^-$, $5/2^+$, $7/2^+$ and $2^+$,
but poor for $1/2^-$ and $1^+$ hadrons.  It has been speculated that
the mixing between multiplets might be useful for the interpretation
to assign these particles to SU(3) multiplets~\cite{GT69}.

 Recently, Guzey and Polyakov reviewed the spectrum of all baryons with mass
less than approximately 2000-2200~MeV and catalogued them into
twenty-one SU(3) multiplets~\cite{GP}. That work can be viewed as an
attempt to update Ref.~\cite{SGM}. In their paper, the masses for
(N, $\Lambda$, $\Sigma$, $\Xi$) members of an octet are listed in a
parenthesis. They introduced the mixing of the octet (\textbf{8},
$1/2^-$)=(1535, 1670, 1560, \underline{1620-1725}) with the singlet
(\textbf{1}, $1/2^-$)=$\Lambda$(1405). It is noticed that the small
$\Gamma_{\Lambda(1670) \to N {\overline K}}$
 forces $\Gamma_{N(1535)\to N \pi}$  to be also small, which is in conflict with the
experimental observation.  The mixing of $\Lambda(1405)$ with
$\Lambda(1670)$ enables one to simultaneously have sufficiently
 large $A_8$ and small $g_{\Lambda(1670) \to  N \overline{K}}$. In this way, most
predicted decay widths can fit the experimental data. However,  the
predicted $\Gamma_{N(1535)\to N \eta}$ is lower than
$\Gamma_{N(1535)\to N \eta}$ according to the experimental data form
Particle Data Group (PDG) listing~\cite{PDG}, and the predicted
$\Gamma_{\Sigma(1560)\mathrm{total}}$ is broader than the
experimental value $\Gamma_{\Sigma(1560)\mathrm{total}}$. From
analysis of these situations, Guzey and Polyakov thought that the
practical mixing angle should be smaller than their theoretical
fitted mixing angle. Meanwhile, in the octet (8, $1/2^-$)=(1650,
1800, 1620, \underline{1860-1915}), it will be better if the
predicted $\Gamma_{N(1650)\to N \pi}$ is broader.

In this paper, instead of considering only one octet and one singlet
mixing, we investigate the mixing between the two octets with
$J^p=1/2^-$, which are (1535, 1670, 1560, \underline{1620-1725}) and
(1650, 1800, 1620, \underline{1860-1915}), and then their mixing
with the singlet $\Lambda(1405)$. For the the small
$\Gamma_{\Lambda(1670) \to N \overline{K}}$, we still have to mix
$\Lambda(1405)$ with $\Lambda(1670)$. However, we consider the
effect due to mixing between the two octets. There are two main
reasons for this. One is that particles with same quantum numbers
from different unitary multiplets can mix with each other, the other
is that this can decrease the mixing angle between the octet and the
singlet, and enhance $\Gamma_{N(1535)\to N \eta}$ and
$\Gamma_{N(1650)\to N \pi}$. This renders the theoretical
$\Gamma_{\Sigma(1560)\mathrm{total}}$ in agreement with the
experimental data. Therefore, we try to mix the two octets first,
and then mix the mixed $\Lambda(1670)$ state with $\Lambda(1405)$.

\section{Decay widths and coupling constants}

For the decay process of a baryon B$^*$  to a baryon B and a
pseudoscalar meson M
\begin{equation}B^* \to B + M ,\end{equation}
the calculation of decay widths can be performed in the framework of
Rarita-Schwinger formalism. The parity-conserving Lagrangian of
$B^*_{1/2^-} \to B_{1/2^+} + M $ interaction is~\cite{JGR}
\begin{equation} \mathcal{L}=g_{B^*BM}\bar{\Psi}\Phi\phi, \end{equation}
where $\Psi$ is the $J^p=1/2^-$  field, $\Phi$  is the $J^p=1/2^+$
field, $\phi$ is the pseudoscalar meson field. Meanwhile, the
Lagrangian of $B^*_{1/2^-} \to B_{3/2^+} + M$ interaction is
\begin{equation} \mathcal{L}=i\frac{g_{B^*BM}}{m_{\pi}} \bar{\Psi}\gamma_5\Phi^{\mu}\partial_{\mu}\phi,\end{equation}
where the factor $1/{m_\pi}$ is introduced to make the coupling
constant $g_{B^* BM}$  dimensionless. Accordingly, the decay widths
are written as
\begin{eqnarray}
\Gamma_{B^*_{1/2^-}\to B_{1/2^+}M} &=&  \frac{g^2_{B^{*}BM}}{8\pi m^{*2}_B} P_{cm} [(m^*_B+m_B)^2-m^2],\\
\Gamma_ {B^*_{1/2^-}\to B_{3/2^+}M} &=&  \frac
{g^2_{B^{*}BM}P^3_{cm}[(m^*_B-m_B)^2-m^2]}{24\pi(m^*_Bm_\pi)^2},
\end{eqnarray}
where $g_{B^* BM}$ is the physical coupling constant and
$P_\mathrm{cm}$ is the c.m. momentum of final particles. In terms of
the baryons masses $m_{B}^{*}$, $m_B$ and the meson mass $m$, we
have
\begin{equation}
P_\mathrm{cm}=
\frac{1}{2m^*_B}\{[m^{*2}_B-(m_B+m)^2][m^{*2}_B-(m_B-m)^2]\}^{1/2},
\end{equation}

In order to calculate the partial decay widths, we must know the
physical coupling constant $g_{B^* BM}$, which needs not only
computing the Clebsch-Gordan coefficient among the SU(3) irreducible
representations of B$^*$, B and M, but also computing mixing of two
octets and singlet.

To derive the physical coupling constants,  we first denote the
physical states as $|N_8\rangle$, $|N_{8'}\rangle$,
$|\Lambda_8\rangle$, $|\Lambda_{8'}\rangle$, $|\Lambda_1\rangle$,
$~|\Sigma_8\rangle$, $|\Sigma_{8'}\rangle$, $|\Xi_8\rangle$,
$|\Xi_{8'}\rangle$, and the bare states as $|N^0_8\rangle$,
$|N^0_{8'}\rangle$, $|\Lambda^0_8\rangle$, $|\Lambda^0_{8'}\rangle$,
$|\Lambda^0_1\rangle$, $|\Sigma^0_8\rangle$,
$|\Sigma^0_{8'}\rangle$, $|\Xi^0_8\rangle$, $|\Xi^0_{8'}\rangle$.
The denotations without prime are for the octet (1535, 1670, 1560,
\underline{1620-1725}) and with prime are for the octet (1650, 1800,
1620, \underline{1860-1915}). Then, we introduce the mixing between
the two octets. We assume that the mixing angles of different
baryons in the two octets are different, i.e., the angles are
$\delta$, $\theta$, $\beta$, $\gamma$ for N, $\Lambda$, $\Sigma$,
$\Xi$, respectively. Since the physical states can be written as
linear superpositions of the bare states, the physical states
$|N_8\rangle$, $|N_{8'}\rangle$ are
\begin{equation}
\left(\begin{array}{c}
   |N_8\rangle\\
   |N_{8'}\rangle
   \end{array}\right)=
   \left(\begin{array}{cc}
    \cos\delta & \sin\delta\\
     -\sin\delta &  \cos\delta
     \end{array}\right)\left(\begin{array}{c}
   |N^0_8\rangle\\
   |N^0_{8'}\rangle
   \end{array}\right).
\end{equation}
Therefore, the  coupling constants of Ns after two octet mixing are
\begin{equation}
\left(
   \begin{array}{c}
     g_{N_{8}BM}\\
     g_{N_{8'}BM}
   \end{array}
\right)=
\left(
   \begin{array}{cc}
     \cos\delta & \sin\delta\\
     -\sin\delta &  \cos\delta
   \end{array}
\right)
\left(
   \begin{array}{c}
    g_{N^0_{8}BM}\\
    g_{N^0_{8'}BM}
   \end{array}
\right),\\
\end{equation}
where $g_{B^{*0}BM}$ is the SU(3) universal coupling constant,
$g_{B^*BM}$ is the  physical coupling constant of mixing. The
physical states of  $\Sigma$, $\Xi$ have the same form as N.
However,  $\Lambda$s are special, the  couplings after mixing with
octet are
\begin{eqnarray}
g_{\tilde{\Lambda}(1670)BM}&=&~~\cos\theta~g_{\Lambda^0(1670)BM}+ \sin\theta~ g_{\Lambda^0(1800)BM} ,\nonumber \\
g_{\Lambda(1800)BM}&=&-\sin\theta~ g_{\Lambda^0(1670)BM}+\cos\theta~g_{\Lambda^0(1800)BM},
\end{eqnarray}
where $\widetilde{\Lambda}(1670)$ represents a middle state. And
then, we introduce the angle $\xi$ to mix the mixed state
$\widetilde{\Lambda}{(1670)}$ with $\Lambda{(1405)}$ and  can get
the coupling constants of physical state $\Lambda(1670)$ and
$\Lambda(1405)$
\begin{eqnarray}
g_{\Lambda(1405)BM} & = &\cos\xi~g_{\Lambda^0(1405)BM} + \sin\xi~g_{\tilde{\Lambda}(1670)BM},\nonumber  \\
g_{\Lambda(1670)BM} & = &-\sin\xi~g_{\Lambda^0(1405)BM}+
\cos\xi~g_{\tilde{\Lambda}(1670)BM}.
\end{eqnarray}

 In these physical coupling constants, there are 12 parameters, which are five mixing angle
parameters ($\delta$, $\theta$, $\beta$, $\gamma$, $\xi$) and 7
coupling parameters  ($\alpha$, $A_8$, $A_{10}$, $A_1$, $\alpha'$,
$A'_8$, $A'_{10}$) of the universal coupling constants~\cite{SGM},
where the parameters $\alpha, A_8$, $\alpha', A'_8$ are for the
${\bf 8} \to {\bf 8}+{\bf 8}$ decays, $A_{10}$, $A'_{10}$ for the
${\bf 8} \to {\bf 10}+{\bf 8}$ decays and $A_1$ for the ${\bf 1} \to
{\bf 8}+{\bf 8}$ decays. If these parameters are known, we can get
all the coupling constants and calculate the decay widths. On the
other hand, we can also constrain the parameters if we know
information on the coupling constants.

 Therefore, we use some experimental decay widths of baryons to adjust the
parameters of coupling constants, then use these parameters to
calculate all decay widths. If the calculated decay widths are
consistent with or closer to the experimental data, it supports the
feasibility of considering mixing between octets.

\section{Mass}

We denote the bare masses of baryons as $N^0_8$, $N^0_{8'}$,
$\Sigma^0_8$, $\Sigma^0_{8'}$, $\Xi^0_8$, $\Xi^0_{8'}$,
$\Lambda^0_1$, $\Lambda^0_8$, $\Lambda^0_{8'}$ and the physical
masses as $N_8$, $N_{8'}$, $\Sigma_8$, $\Sigma_{8'}$, $\Xi_8$,
$\Xi_{8'}$, $\Lambda_1$, $\Lambda_8$, $\Lambda_{8'} $. The bare
masses of the two octets are~\cite{GT69}
\begin{equation}\label{eq:GMO}
\begin{array}{ccc}
N^0_8&=&M_1-x_1+y_1,\\
\Lambda^0_8&=&M_1-2x_1,\\
\Sigma^0_8&=&M_1+2x_1,\\
\Xi^0_8&=&M_1-x_1-y_1,
\end{array}\\\hspace{50pt}
\begin{array}{ccc}
 N^0_{8'}&=&M_2-x_2+y_2,\\
\Lambda^0_{8'}&=& M_2-2x_2,\\
\Sigma^0_{8'}&=&M_2+2x_2,\\
\Xi^0_{8'}&=&M_2-x_2-y_2.
\end{array}
\end{equation}
A consequence of Eq.~$(\ref{eq:GMO})$ is
the Gell-Mann--Okubo (GMO) relation for octet masses
\begin{equation}
\frac{N_8+\Xi_8}{2} = \frac{3\Lambda_8+\Sigma_8}{4}.
\end{equation}
The physical masses can be obtained from the diagonalization of the
matrices
\begin{equation}\label{eq:massnomix}
\left(\begin{array}{cc} N^0_8 & V_N\\V_N & N^0_{8'}\end{array}
\right), ~~\left(\begin{array}{cc} \Sigma^0_8 & V_\Sigma\\V_\Sigma &
\Sigma^0_{8'}\end{array}\right), ~~\left(\begin{array}{cc}  \Xi^0_8&
V_\Xi\\V_\Xi &
\Xi^0_{8'}\end{array}\right),~~\left(\begin{array}{ccc}\Lambda^0_1 &
V_1 &V_2\\V_1 &  \Lambda^0_8 & V_3\\V_2 & V_3
&\Lambda^0_{8'}\end{array}\right).\end{equation}

Take the special $\Lambda$ for example, which mixes twice. The first
is the $\Lambda^0_8$ and $\Lambda^0_{8'}$ mixing, the second is the
$\Lambda^0_1$ and $\tilde{\Lambda}_8$ mixing
\begin{equation}
\left(\begin{array}{c}
   |\widetilde{\Lambda}_8\rangle\\
   |\Lambda_{8'}\rangle
   \end{array}\right)=
   \left(\begin{array}{cc}
    \cos\theta & \sin\theta\\
     -\sin\theta &  \cos\theta
     \end{array}\right)\left(\begin{array}{c}
   |\Lambda^0_8\rangle\\
   |\Lambda^0_{8'}\rangle
   \end{array}\right),\nonumber
\end{equation}
\begin{equation}
\left(\begin{array}{c}
   |\Lambda_1\rangle\\
   |\Lambda_{8}\rangle
   \end{array}\right)=
   \left(\begin{array}{cc}
    \cos\xi & \sin\xi\\
     -\sin\xi &  \cos\xi
   \end{array}\right)\left(\begin{array}{c}
   |\Lambda^0_1\rangle\\
   |\widetilde{\Lambda}_8\rangle
   \end{array}\right).
\end{equation}
For simplicity, we use $\Lambda$ instead of  $|\Lambda\rangle$ to
denote the state. We write the bare states from the physical states using the mixing of 3$\times$3 matrices as follows: 
\begin{eqnarray}\label{eq:mixangle}
\left(\begin{array}{c} \Lambda^0_1\\ \Lambda^0_8\\
\Lambda^0_{8'}\end{array}\right)
   &=& \left(\begin{array}{ccc}1 & 0 & 0\\ 0 & \cos\theta & -\sin\theta\\ 0 & \sin\theta &  \cos\theta\end{array}\right)\left(\begin{array}{c}
        \Lambda^0_1\\ \widetilde{\Lambda}_8\\\Lambda_{8'} \end{array}\right) \nonumber\\
   &=&  \left(\begin{array}{ccc}1 & 0 & 0\\
    0 & \cos\theta & -\sin\theta\\
    0 & \sin\theta &  \cos\theta
   \end{array}\right)\left(\begin{array}{ccc}
    \cos\xi & -\sin\xi & 0\\
     \sin\xi &  \cos\xi & 0\\0& 0&1
   \end{array}\right)\left(\begin{array}{c}\Lambda_1\\
   \Lambda_{8}\\
   \Lambda_{8'}\\
   \end{array}\right).
\end{eqnarray}
Using~(\ref{eq:massnomix}) and~(\ref{eq:mixangle}), we can get the
result
\begin{eqnarray}
\hspace{-15pt}\left(
  \begin{array}{ccc}
   \Lambda^0_1 & \Lambda^0_8 &
   \Lambda^0_{8'}
  \end{array}
\right)\left(\begin{array}{ccc}\Lambda^0_1 &
  V_1 &V_2\\V_1 &  \Lambda^0_8 & V_3\\V_2 & V_3
  &\Lambda^0_{8'}\end{array}\right)\left(\begin{array}{c} \Lambda^0_1\\
  \Lambda^0_8 \\\Lambda^0_{8'}\\\end{array}\right)=
&& \left(\begin{array}{ccc} \Lambda_1 & \Lambda_{8} &
\Lambda_{8'}\end{array}\right)\left(\begin{array}{ccc}\Lambda_1 & 0
&0\\0 & \Lambda_{8} & 0\\0 & 0
&\Lambda_{8'}\end{array}\right)\left(\begin{array}{c} \Lambda_1
\\\Lambda_{8} \\ \Lambda_{8'}\\\end{array}\right),
\end{eqnarray}
where $\Lambda$ physical masses are
\begin{eqnarray}
\Lambda_1 &=&
\cos^2\xi\Lambda^0_1+\cos^2\theta\sin^2\xi\Lambda^0_8+\sin^2\theta\sin^2\xi\Lambda^0_{8'}\nonumber\\
 &&+\cos\theta\sin2\xi V_1+\sin\theta\sin2\xi V_2+\sin2\theta\sin^2\xi V_3,\nonumber\\
\Lambda_8 &=&
\sin^2\xi\Lambda^0_1+\cos^2\theta\cos^2\xi\Lambda^0_8+\sin^2\theta\cos^2\xi\Lambda^0_{8'}\nonumber\\
 &&-\cos\theta\sin2\xi V_1-\sin\theta\sin2\xi V_2+\sin2\theta\cos^2\xi V_3,\nonumber\\
\Lambda_{8'}
&=&\sin^2\theta\Lambda^0_8+\cos^2\theta\Lambda^0_{8'}-\sin2\theta
V_3.\label{eq:lambdamass}
\end{eqnarray}
Meanwhile the physical masses of other baryons are~\cite{DP03}
\begin{eqnarray}
 N_{8,8'}&=&\frac{1}{2}\left(
 N^0_8+N^0_{8'}\mp\sqrt{(N^0_8-N^0_{8'})^2+4V^2_N}\right),\nonumber\\
\Sigma_{8,8'}&=&\frac{1}{2}\left(
\Sigma^0_8+\Sigma^0_{8'}\mp\sqrt{~~(\Sigma^0_8-\Sigma^0_{8'})^2+4V^2_\Sigma}\right),\nonumber\\
\Xi_{8,8'}&=&\frac{1}{2}\left(
\Xi^0_8+\Xi^0_{8'}\mp\sqrt{~~(\Xi^0_8-\Xi^0_{8'})^2+4V^2_\Xi}\right),\label{eq:ksimass}
\end{eqnarray}
 where
 \begin{eqnarray}
 V_N&=&\frac{1}{2}(N_8-N_{8'})\sin2\delta,\nonumber\\
 V_\Sigma&=&\frac{1}{2}(\Sigma_8-\Sigma_{8'})\sin2\beta,\nonumber\\
 V_\Xi&=&\frac{1}{2}(\Xi_8-\Xi_{8'})\sin2\gamma,\nonumber\\
 V_1&=&\frac{1}{2}\cos\theta \sin2\xi(\Lambda_1-\Lambda_8),\nonumber\\
 V_2&=&\frac{1}{2}\sin\theta \sin2\xi(\Lambda_1-\Lambda_8),\nonumber\\
 V_3&=&\frac{1}{2}\sin2\xi\left(\sin^2\xi\Lambda_1+\cos^2\xi\Lambda_8-\sin2\theta\Lambda_{8'}\right).
 \end{eqnarray}
From~(\ref{eq:lambdamass}),~(\ref{eq:ksimass}), we get a new
relation among the baryon masses
\begin{eqnarray}
3(\sin^2\xi\Lambda_1+\cos^2\xi\Lambda_8+\Lambda_{8'})+(\Sigma_8+\Sigma_{8'})
&=& 2(N_8+N_{8'}+\Xi_8+\Xi_{8'}).\label{eq:newrelation}
\end{eqnarray}
For getting more information of $\Xi$, we need to know the
parameters. However, there are 12 parameters ($M_1$, $x_1$, $y_1$,
$M_2$, $x_2$, $y_2$, $M_{\Lambda_1}^0$, $\delta$, $\theta$, $\xi$,
$\beta$, $\gamma$), while we only know 7 physical masses ($N_8$,
$N_{8'}$, $\Sigma_8$, $\Sigma_{8'}$, $\Lambda_1$, $\Lambda_8$,
$\Lambda_{8'}$) from PDG~\cite{PDG}. Therefore we use the angles
obtained from decay widths and physical masses to calculate the
parameters $M_1$, $x_1$, $y_1$, $M_2$, $x_2$, $y_2$,
$M_{\Lambda_1}^0$. Then, we can get all bare masses.

Meanwhile, there is a relation
\begin{eqnarray}\label{eq:ksi}
 2V_\Xi &=&(\Xi^0_8-\Xi^0_{8'})\tan2\gamma.
\end{eqnarray}
Using~(\ref{eq:ksimass}),~(\ref{eq:newrelation}),~(\ref{eq:ksi}) and
according to the experimental $\Gamma_{\Xi \mathrm{total}}$, we
predict the mixing angle ranges and the physical mass ranges of the
two $\Xi$'s.

\section{Results}

The detailed calculating processes are as follows. At first, we use
the N and $\Lambda$ experimental decay width data and the least
square method to get the parameters ($\delta$, $\theta$, $\xi$,
$\alpha$, $A_8$, $A_{10}$, $A_1$, $\alpha'$, $A'_8$, $A'_{10}$).
Secondly, we use $\Sigma$ width data to get $\beta$. Thirdly, we use
the angles obtained from decay widths and physical masses to
calculate the parameters $M_1$, $x_1$, $y_1$, $M_2$, $x_2$, $y_2$,
$M_{\Lambda_1}^0$. Finally, we use the obtained angles and the mass
parameters to get $\gamma$ range and the $\Xi$ mass ranges of the
two octets, and further, to predict the decay width ranges of the
two $\Xi$s. The parameters are
\begin{equation}
\begin{array}{ccc}
\delta & =& -4.5 ^\circ,\\
\alpha & =& -0.697,\\
\alpha '& =&0.69,
\end{array}\quad
\begin{array}{ccc}
\theta & =& -26.4^\circ,\\
 A_8 & =& 0.889,\\
 A'_8 & =& 0.69,\\
\end{array}\quad
\begin{array}{ccc}
\xi & =& -34.9^\circ,\\
 A_{10}& =&2.04,\\
 A'_{10}&=&1.05,\\
\end{array}\quad
\begin{array}{ccc}
\beta & =& -31.4^\circ,\\
A_1 & =&1.65,\\
 &&\\
\end{array}
\end{equation}
and%
\begin{equation}
\begin{array}{ccc}
M_1 & =&1601.2,\\
M_2 & =& 1680.4,\\
\end{array}\quad
\begin{array}{ccc}
x_1 & =&-12.5,\\
x_2 & =&-38.4,\\
\end{array}\qquad
\begin{array}{ccc}
y_1& =& -77.9,  \\
 y_2 & =&-69.5. \\
\end{array}\quad
\begin{array}{ccc}
M^0_{\Lambda1}& = &1491.7,\\
&&\\
\end{array}
\end{equation}
According to the experimental $\Gamma_{\Xi \mathrm{total}}$, we
predict that the $\Xi$ mixing angle range $\gamma$ is between
$-36.0^\circ$ and $-29.0 ^\circ$, meanwhile, we predict that two
possible $\Xi$ mass ranges of the two octets are
$\Xi$(\underline{1583-1649}) and $\Xi'$(\underline{1831-1896}). We
list all results in {\it Table 1.} Most predicted decay widths are
consistent with the experimental data.

\begin{table}[h]
\tbl{The masses and widths of baryons (in unit of MeV).}
{\begin{tabular}{@{}cccccc@{}} \toprule PDG & Width  & Decay mode & Branching ratio &$\Gamma_i(\mathrm{exp})$  & $\Gamma_i(\mathrm{th})$\\
\colrule
$N(1535)$ & 125-175 & $N\pi$ &  35\%-55\% & 43.75-96.25 & 96.03\\
& & $N\eta$ & 45\%-60\% & 56.25-105.0 &  57.62 \\
& & $\Delta\pi$ &$<$1\% &  $<$1.75 & 1.73\\
$\Lambda(1670)$ & 25-50 & $N\overline{K}$ & 20\%-30\% & 5.0-15.0 &  5.19\\
& & $\Lambda\eta$ & 10\%-35\% & 2.5-17.5 & 7.21 \\
& & $\Sigma\pi$ & 25\%-55\% & 6.25-27.5 & 13.20\\
& & $\sqrt{\Gamma_{N\overline{K}}\Gamma_{\Sigma(1385)\pi}}$
&$-23$\%- $-11$\% &
$-11.5$- $-2.75$ & $-0.81$\\
$\Lambda(1405)$ & 48-52 & $\Sigma\pi$ & 100\% & 48.0-52.0 & 51.77\\
$\Sigma(1560)$ & 49-109 & $\Gamma_{\Sigma\pi}/(\Gamma_{\Sigma\pi}+\Gamma_{\Lambda\pi})$ &  & 0.23-0.47 & 0.47\\
& & $\Gamma_{\Sigma\pi}$ & & &12.57\\
& & $\Gamma_{\Lambda\pi}$ & & &14.15\\
& & $\Gamma_{N\overline{K}}$ & & &29.02\\
$\Xi(\underline{1583\textendash1649})$ & & & & &\\
& & $\Gamma_{\Xi\pi}$ & & &22.00-70.00\\
& & $\Gamma_{\Lambda\overline{K}}$ & & & 3.00-10.00\\
\colrule
$N(1650)$ & 145-185 & $N\pi$ &  60\%-95\% & 87.0-175.75 &88.10\\
& & $N\eta$ & 3\%-10\% & 4.35-18.5 & 18.47 \\
& & $\Lambda K$ & 3\%-11\% &  4.35-20.35  & 18.95 \\
& & $\Delta\pi$ &1\%-7\% &  1.45-12.95  & 3.34   \\
$\Lambda(1800)$ & 200-400 & $N\overline{K}$ & 25\%-40\% & 50.0-160.0 & 151.93\\
& & $\sqrt{\Gamma_{N\overline{K}}\Gamma_{\Sigma\pi}}$ &$-13$\%-
$-3$\% &
$-52.0$- $-6.0$ & $-51.87$\\
& & $\sqrt{\Gamma_{N\overline{K}}\Gamma_{\Sigma(1385)\pi}}$
&2.8\%-8.4\%
&5.6-33.6 & 22.28\\
$\Sigma(1620)$ & 45-85 & $\Gamma_{N\overline{K}}$ & 20\%-24\% & 9.0-20.4 & 9.02      \\
& & $\sqrt{\Gamma_{N\overline{K}}\Gamma_{\Sigma\pi}}$ & 34\%-46\%&15.3-39.1 &18.01\\
& & $\Gamma_{\Sigma\pi}$ & & &35.98\\
$\Xi(\underline{1831\textendash 1896})$ & & & & &\\
& & $\Gamma_{\Xi\pi}$ & & & 18.00-19.00\\
& & $\Gamma_{\Lambda\overline{K} }$ & & & 34.00-37.00\\
& & $\Gamma_{\Sigma\overline{K} }$ & & & 81.00-86.00\\ \botrule
\end{tabular}}
\end{table}

\section{Discussions}
 In our results, most predicted widths are consistent with the experimental data.
The predicted $\Gamma_{N(1535)\to N \eta}$ and $\Gamma_{N(1650)\to N
\pi}$ become broader, which fit the experimental data.  The
theoretical $\Gamma_{\Sigma(1560)\textrm{total}}$ agrees with the
experimental data. The  mixing angle between the octet and the
singlet is $-34.9^\circ$, which is smaller and in the range
$15^\circ<|\xi|< 35^\circ$~\cite{GP}. It also predicts two possible
$\Xi$ mass ranges. However, the
$\sqrt{\Gamma_{N\overline{K}}\Gamma_{\Sigma(1385) \pi}}$ is lower.
There are three reasons for this. One is that the coupling and the
phase space of $\Gamma_{N(1535)\to \Delta \pi}$ are both bigger than
that of $\Gamma_{N(1670)\to \Sigma(1385) \pi}$; the second is that
the small $\Gamma_{N(1535)\to \Delta \pi}$(most 1.75~MeV) makes
$A_{10}$ not large; and the third is that the mixing of two octets
suppresses $\Gamma_{N(1670)\to \Sigma(1385) \pi}$ by a factor 2.03.

We also have tried three possible ways to include the mixing. The
first is just what we have done above, but to introduce same octet's
mixing angles before their mixing with the singlet, the second is to
consider only two octet mixing without their mixing with the
singlet, the third is to consider the mixing of the  octet with the
singlet firstly and then consider the mixing between the two octets.
However, we cannot find appropriate results which are consistent
with all experimental data. So as an example, we only show the
results from the first way, i.e., to consider the two octet mixing
first, and then mix the mixed state $\widetilde{\Lambda}(1670)$ with
$\Lambda(1405)$. The results suggest that the octet mixing might be
a feasible effect to improve the agreement between theory and
experiments. However, it also suggests us that other mechanism
and/or dynamical effects need to be introduced for a better
description of all available experimental data.

\section*{Acknowledgments}

We are grateful to Qihua Zhou and Bin Wu for useful discussions.
This work is partially supported by National Natural Science
Foundation of China (Nos.~10421503, 10575003, 10528510), by the Key
Grant Project of Chinese Ministry of Education (No.~305001), by the
Research Fund for the Doctoral Program of Higher Education (China).

\end{document}